\documentclass[conference]{IEEEtran}
\usepackage{graphicx} 
\usepackage{tabularx}
\usepackage{float} 
\usepackage{amsmath}
\usepackage{booktabs}
\usepackage{longtable}
\usepackage{multirow}
\usepackage{algorithm}
\usepackage{algpseudocode}
\usepackage{cancel}
\usepackage{subcaption}
\usepackage{multicol}
\usepackage{url}

\usepackage{hyperref}
\hypersetup{
    colorlinks=true,
    linkcolor=black,
    filecolor=magenta,
    citecolor=black,
    urlcolor=blue,
    }

\usepackage{colortbl, xcolor}
\usepackage[most]{tcolorbox}

\usepackage{listings}
\usepackage{tikz}
\usetikzlibrary{calc}
\usepackage{array}

\newcommand{\sectopic}[1]{\vspace*{0.1em}\par\noindent{\textit{\bfseries #1}}}

\title{Test Input Validation for Vision-based DL Systems: An Active Learning Approach}

\author{
	\IEEEauthorblockN{Delaram Ghobari\IEEEauthorrefmark{1},
                      Mohammad Hossein Amini\IEEEauthorrefmark{1},
                       Dai Quoc Tran\IEEEauthorrefmark{2},
                      Seunghee Park\IEEEauthorrefmark{2}, \\
                      Shiva Nejati\IEEEauthorrefmark{1},
                      Mehrdad Sabetzadeh\IEEEauthorrefmark{1}}

    \IEEEauthorblockA{\IEEEauthorrefmark{1}University of Ottawa, Ottawa, Canada}
    \IEEEauthorblockA{\IEEEauthorrefmark{2}SmartInsideAI Company Ltd. and Sungkyunkwan University, Suwon, South Korea}
	Email: \{dghob088, mh.amini, snejati, m.sabetzadeh\}@uottawa.ca; 
 \{daitran, shpark\}@smartinside.ai
}

\newcommand{\metrics}{image-comparison metrics}

\newcommand{\approach}[0]{HiL-TV}

\begin{document}

\maketitle
\thispagestyle{plain}

\begin{abstract}
Testing deep learning (DL) systems requires extensive and diverse, yet valid, test inputs. While synthetic test input generation methods, such as metamorphic testing, are widely used for DL testing, they risk introducing invalid inputs that do not accurately reflect real-world scenarios. Invalid test inputs can lead to misleading results. Hence, there is a need for automated validation of test inputs to ensure effective assessment of DL systems. In this paper, we propose a test input validation approach for vision-based DL systems. Our approach uses active learning to balance the trade-off between accuracy and the manual effort required for test input validation. Further, by employing multiple image-comparison metrics, it achieves better results in classifying valid and invalid test inputs compared to methods that rely on single metrics. We evaluate our approach using an industrial and a public-domain dataset. Our evaluation shows that our multi-metric, active learning-based  approach produces several optimal accuracy-effort trade-offs, including those deemed practical and desirable by our industry partner. Furthermore, provided with the same level of manual effort, our approach is significantly more accurate than two state-of-the-art test input validation methods, achieving an average accuracy of 97\%. Specifically, the use of multiple metrics, rather than a single metric, results in an average improvement of at least 5.4\% in overall accuracy compared to the state-of-the-art baselines. Incorporating an active learning loop for test input validation yields an additional 7.5\% improvement in average accuracy,  bringing the overall average improvement of our approach to at least 12.9\% compared to the baselines.
\end{abstract}
\section{Introduction}
\label{sec:intro}
Deep learning (DL) systems are now a key part of many advanced applications. They can automate complex tasks like anomaly detection, object recognition, and semantic segmentation. Since DL systems rely heavily on data, their effective testing and verification depend on the availability of extensive and diverse test inputs. To address this need, various synthetic test input generation methods have been proposed~\cite{Pei_2019_Deepxplore, Dola_2021_DAIV, marsha, WhenWhyTest2022, Tian_2018_deeptest, Zhang2018Deeproad}. These include metamorphic testing approaches, which create new test inputs by systematically modifying existing ones while preserving certain properties. Various test input generation techniques for vision-based DL systems aim to create diverse test inputs by applying transformations such as rotation or scaling~\cite{WhenWhyTest2022, Pei_2019_Deepxplore, Guo_2018_DLFuzz}.

Transforming or synthetically generating test inputs may lead to \emph{invalid} ones. According to recent research in the software testing literature~\cite{WhenWhyTest2022, Dola_2021_DAIV, Stocco_2020_selforacle}, valid test inputs are those that fall within the expected distribution and satisfy the constraints defined by the training data of the DL system under test, while invalid inputs deviate from these expectations and constraints. In particular, Riccio and Tonella~\cite{WhenWhyTest2022} define invalid inputs for a DL system as those that cannot be confidently recognized and labelled by human  experts within the input domain.  Using invalid inputs for testing DL systems can lead to misleading results, 
such as identifying spurious errors -- false or irrelevant errors that would never occur within the system’s intended operational boundaries. Furthermore, invalid inputs may give a false sense of confidence in the system's reliability, as the test time budget is spent assessing the system with inputs it will not encounter in real-world situations, which means that flaws that may arise with valid inputs could go undetected.  Therefore, test inputs need to be validated before being used for testing.

The importance of test input validation for vision-based DL systems is well recognized in software engineering, and techniques have already been developed to address this need~\cite{marsha,WhenWhyTest2022}.  Existing techniques rely on image-comparison metrics, which quantify how well a transformed image aligns with its corresponding original image. Notably, Hu et al.~\cite{marsha} use visual information fidelity (VIF)~\cite{vif}, while Riccio and Tonella~\cite{WhenWhyTest2022} rely on the loss function of a variational autoencoder (VAE)~\cite{Kingma_2014_VAE} trained on a set of valid test inputs.  Both approaches identify an optimal threshold for their respective image-comparison metric, derived from a manually validated set of test inputs. To automatically determine whether a transformation of an original image is valid, the metric value for the pair of the original and transformed image is compared against the identified thresholds. Since the metric values vary considerably from one dataset to another, the thresholds  do not necessarily generalize to datasets not considered in the original papers. 
Further, these approaches use only a single metric. It is unclear whether a single metric can effectively distinguish valid inputs for different datasets, such as industrial and specialized-domain datasets, which, in contrast to the datasets used by these approaches~\cite{marsha,WhenWhyTest2022}, are not limited to common objects such as animals and vehicles.

In this paper, we propose \approach, an automated, human-in-the-loop test input validation approach for vision-based DL systems. \approach\ receives a set of image pairs, each consisting of an original image and its transformed version, and distinguishes the valid pairs --  
those where the transformed image is an acceptable alteration of the original image -- from the invalid ones. \approach\ uses a machine learning classifier that operates on image-comparison metrics as input features rather than raw images to determine whether a transformation is valid or invalid. Initially, \approach\ trains and tunes the classifier using a labelled subset of the dataset. It then utilizes an active learning loop that continuously improves the classifier's  ability to differentiate between valid and invalid pairs. When the classifier's confidence in a prediction falls below a user-defined threshold, the corresponding image pair is flagged for human review. A human annotator subsequently labels the pair as either valid or invalid. The primary advantage of this active learning loop is that it minimizes human effort by requesting intervention only when necessary, specifically for image pairs that the classifier is unable to validate.

We evaluate \approach\ using two datasets: one from industry and the other from the open-source domain. Our industry dataset was developed by our partner, SmartInside AI~\cite{smartinsideAI}, and includes images of power-grid facilities as well as transformations of these images aimed at simulating foggy, rainy, and snowy weather conditions. As for the  public-domain dataset, we use the dataset developed by Hu et al.~\cite{marsha}. Our evaluation shows that our multi-metric, active learning-based approach to test input validation produces several optimal accuracy-effort trade-offs, including those deemed practical and desirable by our industry partner. Further, we compare \approach\ with two baselines from the software engineering literature: one by Hu et al.~\cite{marsha} and the other by Riccio and Tonella~\cite{WhenWhyTest2022}. We show that, when provided with the same level of manual effort, our approach is significantly more accurate than these two state-of-the-art methods, achieving average accuracy, precision, and recall of 97.0\%, 96.7\%, and 99.5\%, respectively. Specifically, the use of multiple metrics, as opposed to a single metric, results in average increases of 4.6\% and 15.0\% in accuracy, and average increases of 5.4\% and 16.0\% in precision, compared to Hu et al.~\cite{marsha} and Riccio and Tonella~\cite{WhenWhyTest2022}, respectively. Moreover, the introduction of an active learning loop for test input validation provides average improvements of 7.5\% in accuracy, 6.8\% in precision, and 2.1\% in recall compared to our approach without active learning. These findings indicate that both novel features of \approach\ -- using multiple image-comparison metrics and applying active learning -- lead to significant improvements in accuracy and precision.

Finally, our correlation analysis shows that multiple distinct image-comparison metrics have a meaningful relationship with how test inputs are validated across both the datasets in our evaluation. This suggests that using multiple metrics together is more effective than relying on just one.

We conclude the paper by presenting key lessons learned based on our empirical results.

\textbf{Novelty.} The novelty of our work is in the development of an active learning loop that enables users to explore trade-offs between validation accuracy and manual labelling effort. Furthermore, we make use of 13 \metrics\ for test input validation, including two that have been used individually for this purpose in earlier research~\cite{marsha, WhenWhyTest2022}.

\textbf{Significance.} To the best of our knowledge, our work is the first empirically documented application of test input validation for vision-based DL systems in an industrial context. Using industry data,  we demonstrate that our approach outperforms two state-of-the-art test validation methods recently proposed by the software engineering community~\cite{marsha, WhenWhyTest2022}.

\begin{table*}[th]
\centering
\caption{Image-comparison metrics: For each metric, the table provides its source and definition, and categorizes it as either pixel-level or feature-level. Feature-level metrics rely on representations derived from the latent space of pre-trained DL models.}~\label{tab:metrics_explained}
\scalebox{1}{\begin{tabular}{|p{2cm}|p{12.5cm}|p{2cm}|}
  \hline
  Metric & Definition & Comparison level\\ 
  \hline
  PSNR \cite{psnr}& 
  Peak signal-to-noise ratio (PSNR) measures the ratio of the maximum possible pixel value to the mean squared error (MSE) between the original and transformed images. 
  &  Pixel-level 
 \\ 
  \hline
  SSIM \cite{ssim}& 
Structural similarity index measure (SSIM) evaluates the quality of a transformed image by comparing its structural information, e.g.,  edges, textures, and patterns, with the original image. 
& Pixel-level
 \\ 
  \hline
  MSE& 
  Mean squared error (MSE) is the average squared difference between pixel values of the transformed and original images. 
  & Pixel-level
   \\ 
  \hline
  TSI \cite{tsi} & 
  Texture similarity index (TSI) measures the texture properties of the original and transformed images using gray-level co-occurrence matrices.
  & Pixel-level
   \\ 
  \hline
  WS \cite{wd} & 
  Wasserstein score (WS) measures the similarity between two images by calculating the minimum effort needed to transform one image into the other. It quantifies the cost of redistributing pixel values from the generated image to match the reference image.
  & Pixel-level
   \\ 
  \hline
  CS \cite{cs} & 
  Cosine similarity (CS) measures the similarity between high-level feature representations of the original and transformed images created by the convolutional layers of a pre-trained deep-learning model using cosine similarity. 
  & Feature-level
   \\ 
  \hline
  KL \cite{kl} & 
  Kullback–Leibler (KL) divergence  measures the difference between the pixel value distribution of the original and transformed images.
  & Pixel-level
  \\ 
  \hline
  $\mathit{Hist}_{\mathit{int}}$ \cite{histi} & 
  Histogram intersection ($\mathit{Hist}_{\mathit{int}}$) measures the similarity by computing the intersection of the pixels histograms of the transformed and original images.
  & Pixel-level
   \\ 
  \hline

  $\mathit{Hist}_\mathit{cor}$ \cite{histi} & 
   Histogram correlation ($\mathit{Hist}_{\mathit{cor}}$) measures the similarity by computing the correlation of the pixels histograms of the transformed and original images. 
  & Pixel-level
   \\ 
  \hline
  
  CPL \cite{cpl} & 
  Classifier perceptual loss (CPL) measures the difference in high-level representations obtained from a pretrained deep-learning model of the transformed and original images.
  & Feature-level
   \\ 
  \hline
  SSS \cite{sss} & 
  Semantic segmentation score (SSS) measures the semantic and structural differences by comparing segmentation outputs of the transformed and original images obtained by a pretrained segmentation model.
  & Feature-level
   \\ 
  \hline
  
  VAE-RE \cite{vae}& 
 Variational autoencoder reconstruction error (VAE-RE) measures how far a transformed image is from the distribution of a set of original images. To compute this measure, a VAE needs to be trained on a set of original images. VAE-RE is the loss function of the trained VAE for the transformed images.
  & Feature-level
   \\ 
  \hline

  VIF \cite{vif} & 
  Visual information fidelity (VIF) evaluates the visual quality of an image by quantifying the information shared between the transformed and original images.
  & Pixel-level
   \\ 
  \hline
\end{tabular}
}
\end{table*}

\section{Industrial Context}
\label{sec:context}
SmartInside AI (SIA) provides DL-based anomaly detection solutions for critical infrastructure, such as power grids, and workspaces like construction sites. These solutions identify anomalies using images captured from various sources, including closed-circuit cameras, drones, and ground personnel. SIA serves clients across different geographical regions, each with its own specific climatic conditions. One of the company's main priorities is to ensure that its solutions maintain accuracy across its clients' diverse environments.

\begin{figure}[t]
\centering
\includegraphics[width=\linewidth]{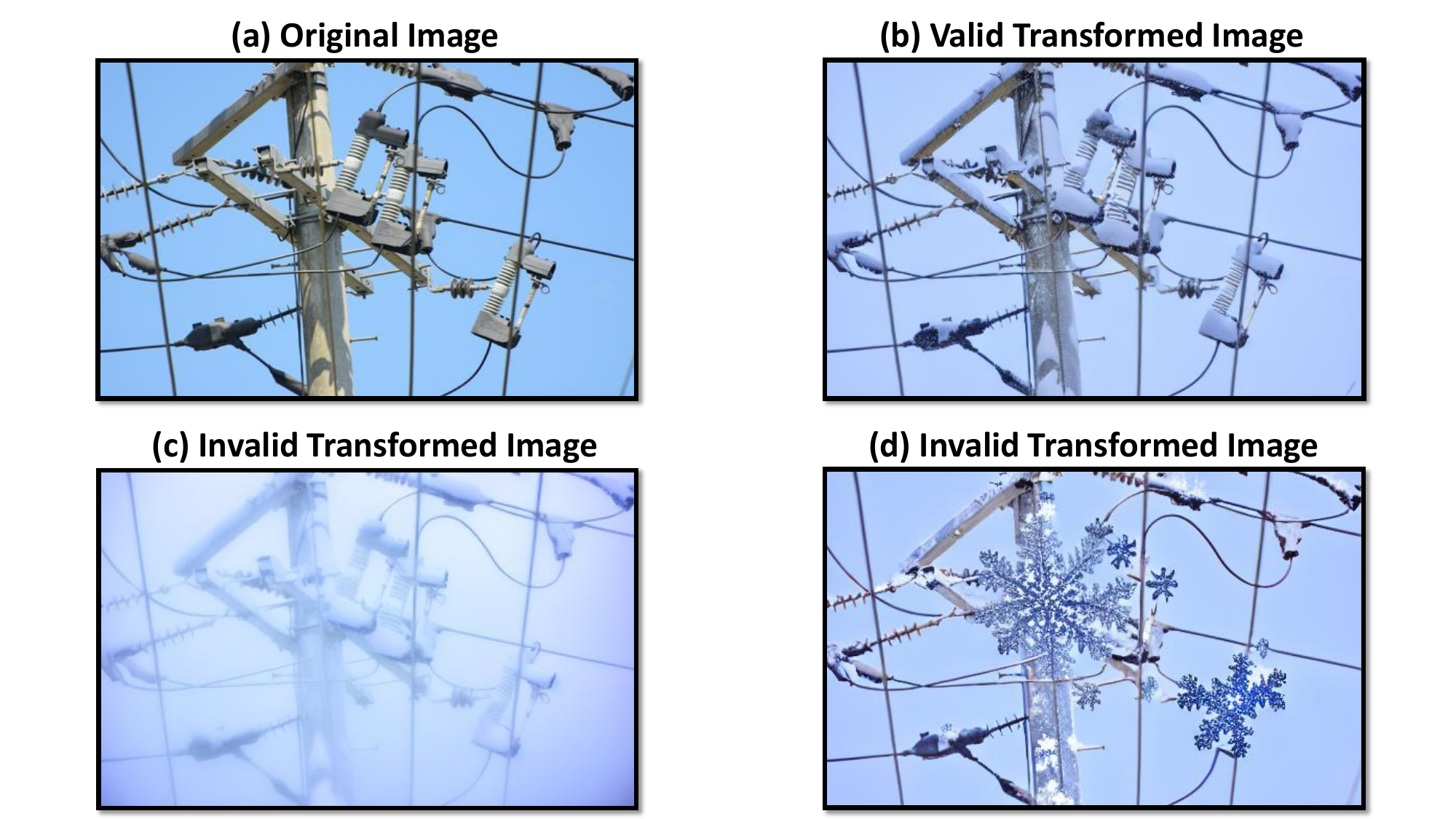}
\caption{An original power-grid image (a) and its transformations to  replicate snowy conditions: (b) is a valid transformation of (a), while (c) and (d) are invalid \hbox{transformations of (a).}}
\label{fig:introexample}
\vspace*{-.3cm}
\end{figure}

To achieve this objective, SIA systematically tests its solutions before deployment in new environments. A key part of this testing involves using synthetic images that replicate the environmental conditions of the target location, such as snow, dust, fog, mist, or frost. SIA generates these synthetic images via transformation techniques, e.g.,  geometric, colour, and noise adjustments, using Albumentations~\cite{Albumentations2020} and  generative techniques such as GANs~\cite{Goodfellow2014GAN} and diffusion models~\cite{Brooks2023Instructpix2pix}. The generated images are then used for robustness testing.

A significant challenge in this context is that some synthetic images are invalid due to excessive distortion or unrealistic features, making them unsuitable for  testing. For example, consider Figure~\ref{fig:introexample}. Figure~\ref{fig:introexample}(a) is an image originally taken in South Korea, which SIA aims to modify to simulate the snowy conditions typical of Canada. Figures~\ref{fig:introexample}(b)–(d) show three images synthesized with snowy conditions. Figure~\ref{fig:introexample}(b) would be deemed valid by experts for anomaly detection purposes. In contrast, Figure~\ref{fig:introexample}(c) is invalid because the white-out conditions and low visibility would make it impossible, even for experts, to detect anomalies through direct visual inspection. Finally, Figure~\ref{fig:introexample}(d) is invalid due to the nonsensical addition of snowflakes into the image.

Identifying and filtering out invalid images manually is a time-consuming process. Our goal  is to devise an approach based on active learning to improve the efficiency of this validation process, thereby reducing the manual effort involved.
\section{Test Input Validation via Active Learning}
\label{sec:approach}
In this section, we introduce \approach, our approach for validating test inputs in vision-based DL systems. The core of \approach\ is a classifier that differentiates between valid and invalid image pairs. As mentioned in Section~\ref{sec:intro}, an image pair consists of an original image and a transformation thereof. An image pair is considered valid if and only if the transformed image represents a legitimate modification of the original. Given a dataset of image pairs, along with a confidence threshold for accepting classifier predictions, \approach\ determines the validity of the pairs by incorporating human input as needed to meet the specified confidence level. To train an effective classifier for its active learning loop, \approach\ requires a subset of the image pairs in the dataset to be pre-validated -- that is, labelled as valid or invalid beforehand.

The classifier component of \approach\ uses as input features the image-comparison metrics listed in Table~\ref{tab:metrics_explained}. These metrics, identified through our review of the literature on methods for comparing image content and its transformations~\cite{marsha, WhenWhyTest2022, stocco}, quantify the similarity or difference between the original and transformed images, helping to determine whether the essential content is preserved. For each metric in Table~\ref{tab:metrics_explained}, we include its source, provide a brief definition, and specify the level at which it compares images. The pixel-level metrics in Table~\ref{tab:metrics_explained} compare images directly using pixel values or statistical information from pixel values, whereas the feature-level metrics compare images by extracting high-level representations from the latent space of pre-trained DL models~\cite{Razavina_2014_CNNFeatures}.

We choose not to feed images directly into a classifier to avoid limitations such as high dimensionality and sensitivity to irrelevant variations (e.g., shifts, lighting, noise) that do not indicate meaningful content differences~\cite{vif}. Instead, by using the metrics in Table~\ref{tab:metrics_explained}, we aim to provide a more accurate and robust indicator of whether the essential content is preserved during transformations.
Among the metrics in Table~\ref{tab:metrics_explained}, to our knowledge, only VAE-RE and VIF have been previously used for assessing the validity of transformed test inputs~\cite{WhenWhyTest2022, marsha}.

To build a classifier over the metrics of Table~\ref{tab:metrics_explained}, we experiment with four different machine learning classification techniques: random forest, decision tree, support vector machines (SVM), and logistic regression~\cite{Bishop2007Pattern}. 
We do not consider DL-based classifiers because our active learning loop re-trains the classifier at each iteration. Since re-training DL models is computationally expensive, they are impractical for an interactive loop that requires \hbox{(near-)\,}instantaneous re-training.

The active learning loop of \approach\ is outlined in Algorithm~\ref{alg:active_learning}. This algorithm takes as input: (a) a set of image pairs, consisting of a pre-validated subset ($D_v$) and a non-validated subset ($D_\mathit{nv}$); (b) a confidence threshold $\alpha$, which specifies the level above which classifier predictions are accepted; and (c) a parameter $\beta$, specifying the number of image pairs to be manually validated in each iteration. The output of the algorithm is a validated dataset, $D_\mathit{val}$. 

The algorithm starts by training a classifier $c$ using $D_v$ (lines\,1-2). During each iteration of the while loop, the algorithm uses $c$ to predict labels, i.e. valid or invalid, for the remaining non-validated image pairs. If the classifier's prediction confidence for an image pair exceeds $\alpha$, the prediction is accepted and the pair is moved from  $\mathit{D}_{\mathit{nv}}$ to $\mathit{D}_{\mathit{val}}$ (lines 4--9). When no more pairs can be confidently validated, the user is prompted to manually label (up to) $\beta$ randomly selected, non-validated pairs (lines 10--11). Subsequently, the user-labelled pairs are moved to  $\mathit{D}_{\mathit{val}}$ (line 12), and  the classifier is re-trained  (line 13). The active learning loop terminates when there are no more image pairs to be validated.

\small
\begin{algorithm}[t]
\caption{Active learning of \approach}
\label{alg:active_learning}
{\footnotesize
\begin{flushleft}
\textbf{Input} $\mathit{D}_v$: Pre-validated subset \\ 
\textbf{Input} $\mathit{D}_{\mathit{nv}}$: Non-validated subset\\
\textbf{Input} $\alpha$: Confidence threshold for acceptable predictions\\
\textbf{Input} $\beta$: Number of pairs to be validated by a human per iteration\\[.5em]
\textbf{Output} $D_{\mathit{val}}$: Validated dataset\\
\end{flushleft}
\begin{algorithmic}[1]
\State $D_{\mathit{val}} = D_v$
\State $c \gets \Call{Train}{D_{\mathit{val}}}$ 
\While{$D_{\mathit{nv}} \neq \emptyset$}
    \For{every  $d \in D_{\mathit{nv}}$}
    \State $\mathit{(prediction, confidence)} \gets \Call{Predict}{c, d}$
    \If{$\mathit{confidence} \geq \alpha$}
    \State Add $d$ to $D_{\mathit{val}}$ and remove it from $D_{\mathit{nv}}$
    \EndIf
    \EndFor
    \State Randomly select $D_{\mathit{tmp}} \subseteq \mathit{D}_{\mathit{nv}}$ s.t. $|D_{\mathit{tmp}}| = \min(\beta, |\mathit{D}_{\mathit{nv}}|)$ 
    \State Prompt human to validate all image pairs in $D_{\mathit{tmp}}$ 
    \State Remove $D_{\mathit{tmp}}$ from $D_{\mathit{nv}}$ and add it to $D_{\mathit{val}} $
    \State $c \gets \Call{Train}{D_{\mathit{val}}}$ 
\EndWhile
\State \Return $D_{\mathit{val}}$
\end{algorithmic}}
\end{algorithm}
\normalsize

Intuitively, the active learning step validates as many image pairs as possible in each iteration based on the threshold $\alpha$, which sets the minimum acceptable prediction confidence. If the classifier cannot produce any further predictions with a confidence $\geq \alpha$, it is re-trained using the set $\mathit{D}_{\mathit{val}}$. Since, in each iteration (except maybe the last), $\mathit{D}_{\mathit{val}}$ contains at least $\beta$ newly user-validated pairs compared to the previous iteration, re-training $c$ with this set has the potential to increase the conclusiveness of the (re-trained) classifier for the yet-to-be-validated pairs in future iterations.

In the following section, we empirically evaluate this algorithm by exploring different classification techniques, different proportions of pre-validated pairs, and different values for $\alpha$ and $\beta$. This investigation highlights the trade-off between minimizing human effort through automated labelling and the risk of errors introduced by automation. Based on our analysis, we provide practical guidelines on how to strike a balance between these competing factors.
\section{Empirical Evaluation}
\label{sec:empirical_eval}
In this section, we evaluate \approach\ using two datasets: an industry dataset provided by our collaborating partner and a public dataset. Our experiments begin with \textbf{RQ1}, which explores the trade-offs between the accuracy of \approach's automated label predictions and the manual effort required for test input validation. Following this, \textbf{RQ2} compares the effectiveness of \approach\ against two state-of-the-art methods from the literature~\cite{marsha,WhenWhyTest2022}, both of which use a single image-comparison metric for test input validation. Finally, \textbf{RQ3} examines how well the metrics listed in Table~\ref{tab:metrics_explained} correlate with the test input validation task, as well as whether any redundancy exists among these metrics in this context.

\textbf{RQ1 (Trade-offs between accuracy and human effort).} \emph{How can we configure \approach\ to balance accuracy and human effort effectively?}
We address RQ1 in three steps. First, we run \approach\ with various configurations to assess how they impact accuracy and effort. Next, we obtain feedback from our industry partner on acceptable trade-off levels between these two factors. Finally, we provide insights into how to achieve these trade-off levels through proper algorithm configuration.

\textbf{RQ2 (Comparison with the state of the art).}  \emph{How does \approach\ compare to state-of-the-art test input validation frameworks?} We compare \approach\ with two state-of-the-art frameworks for test input validation~\cite{marsha,WhenWhyTest2022}, both of which rely on a single image-comparison metric -- VIF in one case and VAE-RE in the other. Like \approach, these baselines require a subset of image pairs to be pre-validated. However, in contrast to \approach, they do not include an active learning loop: After training on the pre-validated subset, they lack a mechanism for human interaction and iterative feedback to enhance the quality of classification. To evaluate these frameworks, we use identical pre-validated training sets and vary the set size to 25\%, 50\%, and 75\% of the overall datasets.

\textbf{RQ3 (Most influential image-comparison metrics).} \emph{Which  metrics in Table~\ref{tab:metrics_explained} are the most influential for test input validation?}
To identify the most important metrics for test input validation, we compute the correlations between each metric in Table~\ref{tab:metrics_explained} and the image validity labels in our two datasets. To assess whether the identified influential metrics provide unique insights for test input validation, we examine potential redundancies by calculating the \hbox{correlations between them.}

\subsection{Implementation}
We have implemented \approach\ in Python 3.11. Our implementation includes a GUI built with Streamlit~\cite{streamlit} to facilitate efficient user interaction. Algorithm~\ref{alg:active_learning} is implemented using Scikit-learn~\cite{sklearn}. Our implementation is publicly available~\cite{tool}.

\subsection{Datasets}
\label{subsec:datasets}
In this section, we present our industry and public datasets. In particular, we discuss the source of the ground-truth labels for each dataset. Note that outside of an evaluation setting, \approach\ requires only a subset of a given dataset to be pre-validated (denoted as $D_v$ in Algorithm \ref{alg:active_learning}). However, for evaluation purposes, ground-truth labels are required for the experimental datasets in their entirety.

\textbf{Industry dataset.} This dataset is based on 847 unique images of power facilities mounted on utility poles in South Korea, provided by our industry partner (SIA). These images are part of the training set used by SIA to develop its DL anomaly detection model, as explained in Section~\ref{sec:context}. To simulate Canada's climatic conditions, SIA has transformed the images using InstructPix2Pix~\cite{Brooks2023Instructpix2pix}, a fine-tuned diffusion model~\cite{Ho_2020_DiffusionModel} that applies localized changes while maintaining the images' overall structure. Specifically, SIA has applied nine transformations representing rain, fog, and snow at three intensity levels: light, moderate, and heavy, resulting in a total of $847 \times 9 = 7,623$  transformed images.

To identify invalid transformed images, we engaged two independent labellers who are not authors of this study. The labellers were given precise guidelines for classifying an image as ``invalid'': An image is considered invalid if it lacks key structural elements of power facilities from the original image, or if it displays significant distortions, blurriness, texture degradation, extreme colour alterations, or unnatural lighting. Furthermore, images with hallucination effects, where the transformation introduces non-existent objects or illogically alters the scene, are to be considered invalid.

To increase the reliability of the labelling process, we developed a tool (included in our replication package) that displays a transformed image alongside its source. If the labeller finds no grounds to deem the transformed image invalid according to the given criteria, they label it as ``ok'', indicating its \emph{provisional} validity. However, as we elaborate shortly, ``ok'' images require further review by a domain expert to conclusively judge their validity.

We conducted about three hours of training sessions to properly calibrate labellers, clarify labelling criteria, and ensure consistent understanding. Interrater agreement was measured using Cohen's kappa ($\kappa$). The $\kappa$ value is 0.69 which suggests substantial agreements~\cite{Cohen_1960_Kappa}. We subsequently discarded from the dataset images with disagreements between the labellers, accounting for approximately $13\%$ of the whole dataset. 

The above labelling process effectively identifies unrealistic images or those that deviate from transformation instructions. However, due to the labellers' lack of domain expertise, they could not conclusively determine if the transformation has accurately preserved defects: if the original image shows a defect, say a cracked insulator, the transformed image should reflect this. Similarly, if no defect is present in the original image, the transformed image should not suggest one.

To verify the validity of transformed images, experts at our partner company reviewed all images labelled ``ok'' by the labellers. Their task was to ensure that defects (or the lack thereof) in an original image were accurately reflected in a transformed image. If an ``ok'' image failed to preserve these details, it was deemed ``invalid''. If no issues were found, the image was confirmed as ``valid''.

In Table~\ref{tab:dataset_info}, we provide summary statistics for our dataset, accounting for the discarded images due to disagreements as well as the validation performed by the experts.

\textbf{Public dataset.} Our public dataset builds on the work of Hu et al.~\cite{marsha}, who transformed and labelled images from the CIFAR-10 dataset~\cite{Cifar10} -- a widely used image classification benchmark consisting of images across ten different classes. Specifically, they selected CIFAR-10 images associated with the \textit{car} class and an equal number of images from all other classes combined (e.g., cat and ship classes), treating them as the \textit{not-car} class. They then applied four safety-related image transformations -- \textit{brightness}, \textit{frost}, \textit{contrast}, and \textit{JPEG compression} -- to both car and not-car images using the Albumentations~\cite{Buslaev_2020_albumentation} library.
The images, both transformed and original, were subsequently labelled as car or not-car through crowdsourcing, with a maximum of 200 milliseconds allocated for labellers to label each image.

Since individual images in Hu et al.'s dataset were labelled by multiple labellers, the dataset contains duplicates. We eliminated these duplicates by determining the car/not-car labels through majority voting. This process resulted in 1,574 unique pairs of original and transformed images, which constitute our public dataset. In our experiments, a transformed image is considered ``valid'' if its human labelling matches the label of its original version; otherwise, the transformed image is deemed ``invalid''.

\begin{table}[t]
\centering
\caption{Statistics for our industry and public  datasets.}
\label{tab:dataset_info}
\scalebox{1}{\begin{tabular}{|c|c|c|c|}
  \hline
  Dataset & Size & Valid Pairs & Image Size (Pixels) \\ 
  \hline
  SmartInside (Industry) & 6493 & 54.2 \% & 512 x 320\\ 
  \hline
  CIFAR10  (Public)  & 1574 & 80.7 \% & 32 x 32\\ 
  \hline
\end{tabular}}
\vspace*{-.3cm}
\end{table}

\subsection{Baselines (B-VIF and B-VAE)}
\label{section:baseline}
We consider as baselines two state-of-the-art test input validation techniques. The first baseline, proposed by Hu et al.~\cite{marsha}, calculates the VIF score for each image pair and classifies the image pair as valid or invalid by comparing its score with a threshold optimized on a manually validated subset of the dataset under validation. We refer to the first baseline as \textbf{B-VIF}. 
The second baseline, proposed by Riccio and Tonella~\cite{WhenWhyTest2022}, trains a VAE on the original images in the dataset, learning their distribution. The VAE generates low reconstruction errors for valid (in-distribution) images and high errors for invalid (out-of-distribution) images. To classify an image pair, the VAE reconstruction error (VAE-RE) of the transformed image is calculated and compared to a threshold, which is optimized on a manually validated subset of the dataset. We refer to this second baseline as \textbf{B-VAE}.

\subsection{Evaluation Metrics}\label{subsec:metrics}
We evaluate \approach\ and our baselines via two metrics (not to be confused with the image-comparison metrics in Table~\ref{tab:metrics_explained}). One metric measures the accuracy of test input validation, and the other measures the manual effort required for labelling:

\vspace*{.15cm}
\sectopic{Accuracy} is defined as the percentage of correctly labelled images in an entire dataset. For \approach, this entire dataset corresponds to $D_{\mathit{val}}$ as returned by Algorithm~\ref{alg:active_learning}.

\vspace*{.15cm}
\sectopic{Human Effort} is defined as the ratio of user-validated images to the size of the entire dataset. Since ground-truth data is available for both of our experimental datasets (see Section~\ref{subsec:datasets}), we \emph{simulate} human input by using the corresponding labels for each image pair from the ground truth. This approach, commonly used in the evaluation of active learning methods~\cite{Chetan_2019_ActiveLearning}, eliminates the need for direct human input. For \approach, human effort is required not only during pre-validation but also within the active learning loop (line 11 of Algorithm~\ref{alg:active_learning}).

\subsection{RQ1 Experiments and Results}
\label{section:RQ1}
We present the results of RQ1 in line with the three steps outlined earlier when we introduced the research question.
\sectopic{1. Exploring Accuracy vs. Human Effort Trade-offs.} To explore the accuracy-effort trade-offs generated by Algorithm~\ref{alg:active_learning}, we execute it using different input configurations as follows. First, for constructing the classifier $c$, we consider four alternative classification techniques: random forest, decision tree, SVM, and logistic regression. For the confidence threshold $\alpha$, we use values of $0.8$, $0.85$, $0.9$, $0.95$, and $0.99$, noting that values below $0.8$ do not adequately filter out low-confidence predictions. The parameter $\beta$, representing the number of pairs manually validated in each iteration, is set to $1\%$, $3\%$, $5\%$, $8\%$, $10\%$, and $15\%$ of the initial non-validated test input set. As for the size of the pre-validated subset $D_v$ in Algorithm~\ref{alg:active_learning}, we consider proportions of $10\%$, $15\%$, $20\%$, $25\%$, $30\%$, and $40\%$ of the full dataset. For each experiment, we randomly select $D_v$ based on the specified proportion, using the remaining data as the non-validated test input set ($D_{\mathit{nv}}$).

At the start of each execution of Algorithm~\ref{alg:active_learning}, we need to train a classifier using the subset $D_v$. We perform five-fold cross-validation on $D_v$ to tune the classifier's hyperparameters. After the initial training is complete, we proceed with the active learning iterations, re-training the classifier at each iteration, as shown in Algorithm~\ref{alg:active_learning}. In total, the number of times  Algorithm~\ref{alg:active_learning}  is executed for RQ1 and for each dataset is: $4\,(\mathit{classifiers}) \times 5\,(\alpha \mathit{\ values}) \times 6\,(\beta \mathit{\ values}) \times 6\,(D_v) = 720$.  For each execution, we measure both the accuracy and the human effort of \approach. To compute \metrics\ that depend on specific DL architectures or hyperparameters, we use the default settings provided in the original papers that introduced each metric, as cited in Table~\ref{tab:metrics_explained}.

\begin{figure}[t]
    \begin{subfigure}{\columnwidth}
      \centering
        \includegraphics[width=1.1\columnwidth]{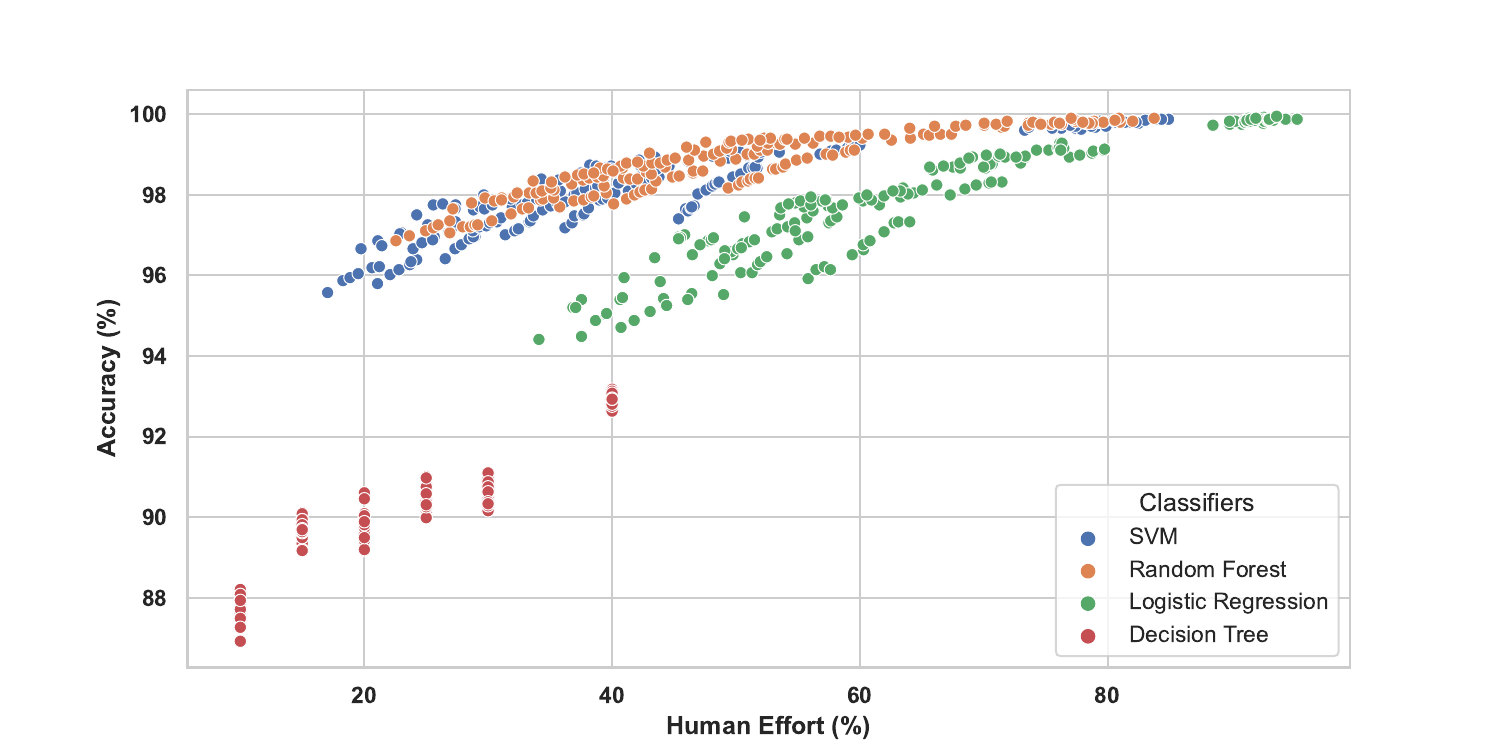}
        \caption{Industry Dataset}
        \label{fig:D1_grid_search}
    \end{subfigure}
    \hfil
    \begin{subfigure}{\columnwidth}
      \centering
        \includegraphics[width=1.1\columnwidth]{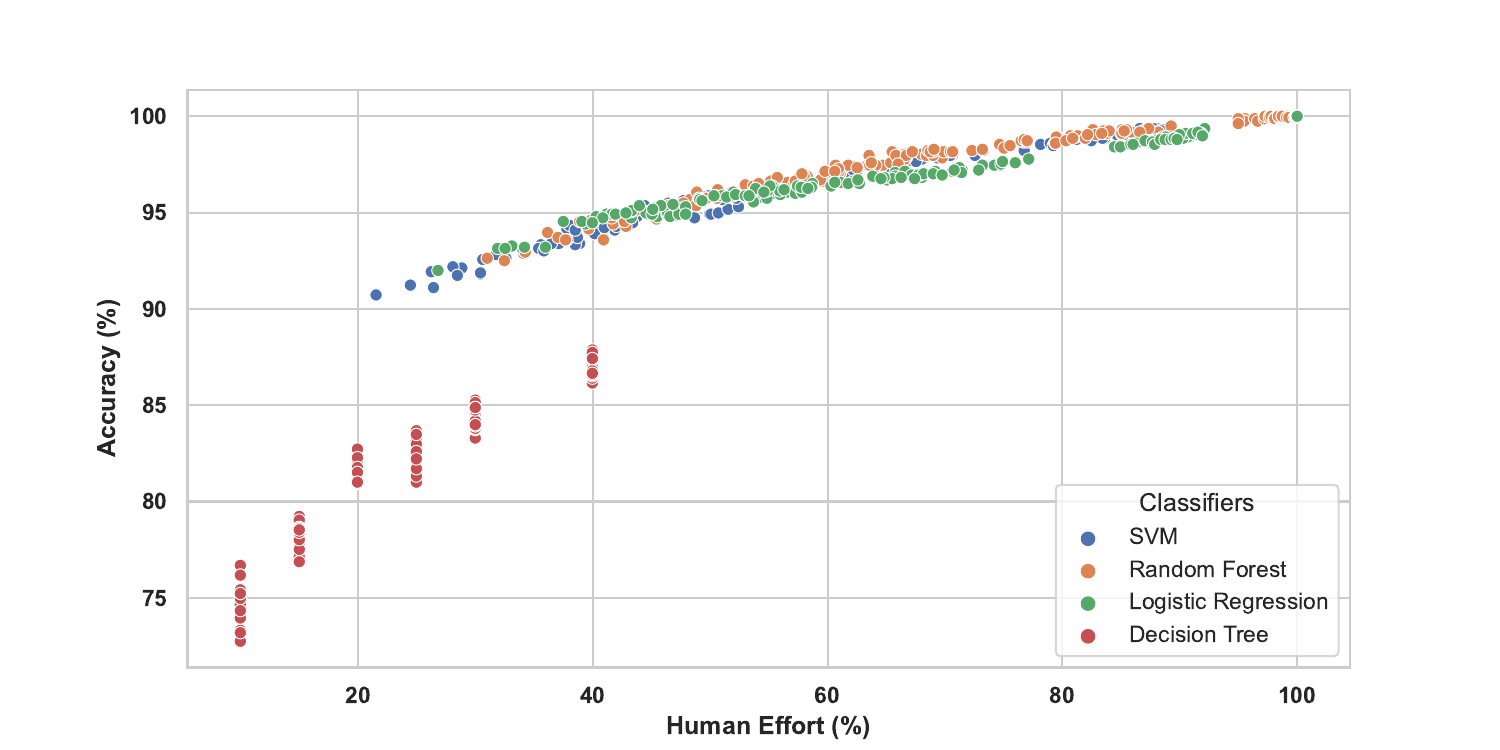}
        \caption{Public Dataset}
        \label{fig:D2_grid_search}
    \end{subfigure}
    \caption{Accuracy vs. human effort trade-offs for various configurations of \approach, based on different classifiers, parameter values of $\alpha$ and $\beta$, and sizes of the pre-validated subset $D_v$.}
    \label{fig:RQ1_gridsearch}
\end{figure}

Figures \ref{fig:RQ1_gridsearch}(a) and (b) show the accuracy vs. human effort for all 720 experiments performed on our industry and public datasets, respectively. Different colours represent the different classification techniques. For the industry dataset, SVM, random forest, and logistic regression achieve an accuracy higher than 95.6\%, 96.9\%, and 94.5\% respectively, regardless of the values of  $\alpha$ and $\beta$ and the size of $D_v$. The decision tree classifier, on the other hand, cannot achieve an accuracy better than 93.2\%. Similarly, decision tree under-performs compared to the other three classifiers for the public dataset, where its accuracy does not exceed 87.9\%. Overall, \approach\ can achieve an accuracy of 99\% with a human effort of 43\% for the industry dataset and 82.1\% for the public dataset, respectively. For the industry dataset, a modest effort of 19.5\% results in a high accuracy of 96\%, while for the public dataset, a similar effort of 19.5\% does not yield an accuracy better than 90.8\%.

Table~\ref{tab:tab_rq1_pareto} shows the number of Pareto points (PPs), representing the optimal trade-offs between accuracy and human effort, for both datasets in Figure~\ref{fig:RQ1_gridsearch}.  A PP is a data point non-dominated by any other point along the two dimensions:  $x$ in the diagrams in Figure~\ref{fig:RQ1_gridsearch} is non-dominated if there is no other point  $y$ that is as good as or better than $x$ in both dimensions, with $y$ being strictly better in at least one dimension. As can be seen from Table~\ref{tab:tab_rq1_pareto}, random forest yields the highest number of optimal trade-offs (PPs) for both datasets.

\begin{table}[t]
\centering
\caption{Number of Pareto points (PPs), representing optimal trade-offs between accuracy and human effort in Figure~\ref{fig:RQ1_gridsearch}, achieved overall and for each classifier: Random Forest (RF), SVM, Logistic Regression (LR), and Decision Tree (DT).}
\label{tab:tab_rq1_pareto}
\scalebox{.9}{
\begin{tabular}{|l|p{1.1cm}|p{1.1cm}|p{1.1cm}|p{1.1cm}|p{1.1cm}|}
\hline
\textbf{Dataset}    & \textbf{\# of PPs (Overall)} & \textbf{\# of PPs by RF} & \textbf{\# of PPs by SVM} & \textbf{\# of PPs by LR} & \textbf{\# of PPs by   DT} \\ \hline
\textbf{Industry} & 57                     & 33                                                                & 16           & 6                                                                       & 2                                                                 \\ \hline
\textbf{Public}     & 128                    & 58                                                                & 17           & 49                                                                      & 4                                                                 \\ \hline
\end{tabular}
}
\vspace*{-.3cm}
\end{table}

\sectopic{2. Industry Partner's Feedback on Trade-off Levels.} Provided with the results in Figure~\ref{fig:RQ1_gridsearch},  our industry partner proposed focusing on two desired accuracy levels for further analysis: 99\% when it can be achieved with reasonable human effort, and 96\% when achieving 99\% would require disproportionate human effort.  Table~\ref{tab:tab_rq3_tradeoff}  presents the details for the optimal trade-offs for our two datasets where accuracy reaches at least 96\% and 99\%. On the industry  dataset, \approach\ achieves 96\% accuracy with 19.5\% human effort using SVM, while reaching 99\% accuracy with 43\% human effort when using a random forest model. For the public dataset, accuracy levels of 96\% and 99\% are achieved using a random forest with human efforts of 48.8\% and 82.1\%, respectively.

\begin{table}[t]
\centering
\caption{Best accuracy-effort trade-offs obtained by \approach, where the accuracy reaches at least 96\% and 99\% (levels indicated to be of interest by our industry partner).}
\label{tab:tab_rq3_tradeoff}
\scalebox{.83}{
\begin{tabular}{|l|l|l|c|c|c|c|}
\hline
\textbf{Dataset}             & \textbf{Classifier} &  $|D_v|$&\textbf{$\alpha$} & \textbf{$\beta$} & \textbf{Accuracy} & \textbf{Human Effort} \\ \hline
\multirow{2}{*}{\textbf{Industry}} & SVM                 &  0.1&0.8& 0.08& 96.0\%& 19.5\%\\ \cline{2-7} 
                             & Random Forest&  0.1&0.95& 0.01          & 99.0\%& 43.0\%\\    \hline
                       
\multirow{2}{*}{\textbf{Public}} & Random Forest&  0.15&0.85& 0.03& 96.0\%& 48.8\%\\ \cline{2-7} 
                             & Random Forest&  0.15&0.95& 0.08& 99.0\%& 82.1\%\\ \hline
\end{tabular}
}
\vspace*{-.5cm}
\end{table}

\begin{figure}[t]
\centering
\includegraphics[width=1\linewidth]{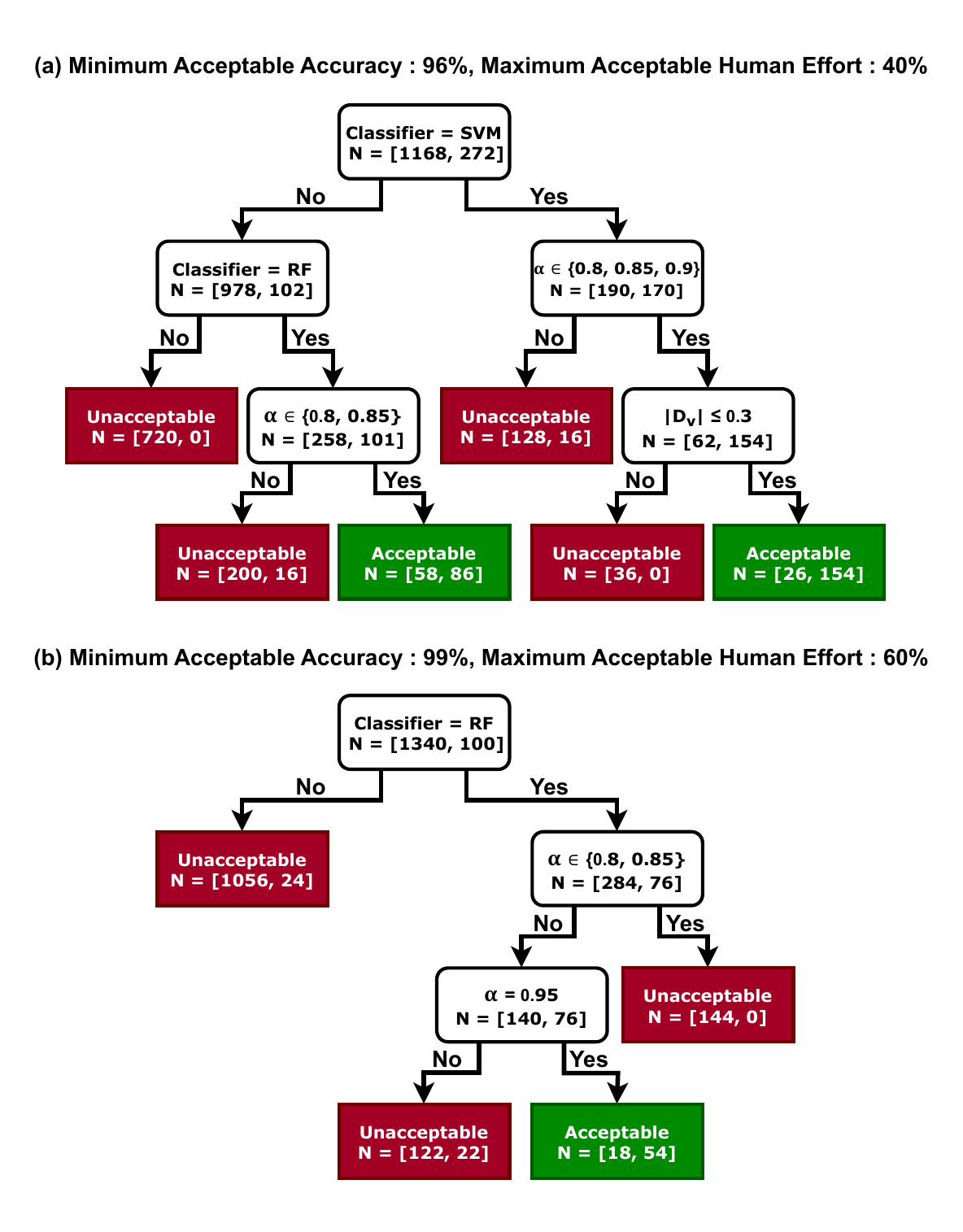}
\caption{Decision trees identifying classification techniques and parameter values that yield acceptable trade-offs between validation accuracy and human effort for our industry partner. At each node, N = [x, y] indicates the count of unacceptable (x) and acceptable (y) configurations reaching that point.}
\label{fig:dt}
\end{figure}

Based on Figure~\ref{fig:RQ1_gridsearch} and Table~\ref{tab:tab_rq3_tradeoff}, our partner decided to cap the human effort budget at 40\% and 60\% for achieving 96\% and 99\% accuracy, respectively. Therefore, the desired accuracy-effort trade-offs are those with an accuracy of at least 96\% when manual effort is at most 40\%, and those with an accuracy of at least 99\% when manual effort is at most 60\%.

\sectopic{3. Achieving the Desired Trade-off Level.} With the desired trade-off levels between accuracy and human effort established above, the final question is which configurations of Algorithm~\ref{alg:active_learning} can achieve these trade-off levels. To address this, we provide a characterization of the acceptable configurations using decision-tree analysis. For this purpose, we combine the data points obtained from both our industry and public datasets, as we want our conclusions to be valid for both, thereby increasing the likelihood of generalizability beyond these specific datasets. 

We partition the data points shown in Figure~\ref{fig:RQ1_gridsearch} (the union of the points in sub-figures (a) and (b)) into acceptable and unacceptable groups for each of the 96\%-40\% and 99\%-60\% trade-off levels. To determine which classification techniques and parameter values yield acceptable outcomes, we construct decision trees over the data points. Figure~\ref{fig:dt} depicts the decision trees for the two trade-off levels we consider. At each node, the tree identifies the most influential parameter for classifying data points as acceptable or unacceptable. As shown in the figure, the most important decision is the choice of classifier. Random forest and SVM are the only classifiers capable of achieving the desired trade-off levels. The second most important decision is the choice of the parameter $\alpha$. An interesting observation from the decision trees is that, within the experimented ranges, $\beta$ and $|D_v|$ values are not important for reaching the desired trade-offs levels. Note that the decision  $|D_v| \leq 0.3$ in Figure~\ref{fig:dt}(a)  trivially follows from capping the manual effort at 40\%. This decision only excludes the value 0.4  among the considered  $|D_v|$ values.

\begin{tcolorbox}[breakable, colback=gray!10!white,colframe=black!75!black]
\textbf{RQ1 Answer:}
Focusing on the two accuracy-effort trade-off levels recommended by our industry partner -- 96\%~vs.~40\% and 99\%~vs.~60\% -- we observe the following: the choice of classifier is the most influential factor. Among the classification techniques considered, only random forest and SVM can achieve the target trade-offs. Random forest yields the greatest number of optimal trade-off points between accuracy and effort, making it the best overall choice. The confidence threshold ($\alpha$) is the second most important factor, generally requiring higher values to filter out low-confidence predictions. The parameters $\beta$ (manual validation per iteration) and $|D_v|$ (pre-validated subset size) have minimal impact within the explored ranges.
\end{tcolorbox}

\subsection{RQ2 Experiments and Results}
\label{section:RQ2}
RQ2 compares \approach\ with the two baselines discussed in Section \ref{section:baseline}. \approach\ differs from these baselines in two key ways: First, \approach\ uses multiple image-comparison metrics, whereas the baselines use only a single (albeit different) metric. Second, \approach\ employs active learning, whereas the baselines do not include any human-in-the-loop components. To assess the impact of these two enhancements, we compare \approach\ with the baselines, both with and without the active learning loop. In these comparisons, we use random forest, which, as established in RQ1, is the best overall choice for classification. As noted in Section~\ref{section:baseline}, we refer to the test input validation method by Hu et al.\cite{marsha} as \textit{B-VIF} and the method by Riccio and Tonella~\cite{WhenWhyTest2022} as \textit{B-VAE}. To account for varying levels of human effort in our comparisons, we set the training set sizes for the compared methods to 25\%, 50\%, and 75\% of the entire dataset.

For the methods that do not rely on active learning, i.e., \textit{B-VIF}, \textit{B-VAE} and \approach\ without active learning, we randomly select a training set from the entire dataset by choosing $x$\% of the dataset where $x \in \{25, 50, 75\}$.   For \approach\ without active learning, we train random forest on the training sets using five-fold cross-validation similar to RQ1. For the \textit{B-VIF} and \textit{B-VAE} baselines, we use the training sets to identify optimal thresholds for the baseline's respective image-comparison metric. Specifically,  for the image pairs in the training sets, we compute VIF values for \textit{B-VIF} and VAE-RE values for \textit{B-VAE}.  Then, for each baseline, we vary its threshold over the range of its respective metric in increments of $10^{-3}$ and calculate the accuracy as the percentage of correctly validated pairs in the training set at each threshold value. The threshold leading to the highest accuracy is selected as the optimal threshold.

For \approach\ with active learning, we execute Algorithm~\ref{alg:active_learning} with random forest  and set $\alpha$,  $\beta$ and $|D_v|$ to the values that yield the lowest manual effort while maintaining a minimum accuracy of 96\% for the industry dataset in RQ1. We use half of the training sets for $D_\mathit{v}$  and cap the number of manual labels during the active learning loop at half the size of the training sets. Once Algorithm~\ref{alg:active_learning} exhausts its manual labelling budget, we terminate the active learning loop and validate the remaining non-validated set using the last random forest classifier trained during active learning. This ensures that Algorithm~\ref{alg:active_learning} does not use more training data than the other methods we compare against.  

To account for randomness, we repeat each experiment 20 times for each method with a random selection of the training set. In total, we run $20 \times 4 \times 3 \times 2 = 480$ experiments for the four methods with three levels of human effort and for our two datasets.
We compare the four methods in terms of \emph{accuracy}, \emph{precision}, and \emph{recall}, using test sets obtained by excluding the training data from the original datasets. Low precision (a high rate of false positives) results in accepting invalid tests as valid, which wastes the testing budget and can trigger unnecessary investigations into misleading failures. On the other hand, low recall (a high rate of false negatives) results in marking valid tests as invalid, leading to the rejection of otherwise good tests.

\begin{figure*}[t]
    \centering
    \includegraphics[width=0.95\textwidth]{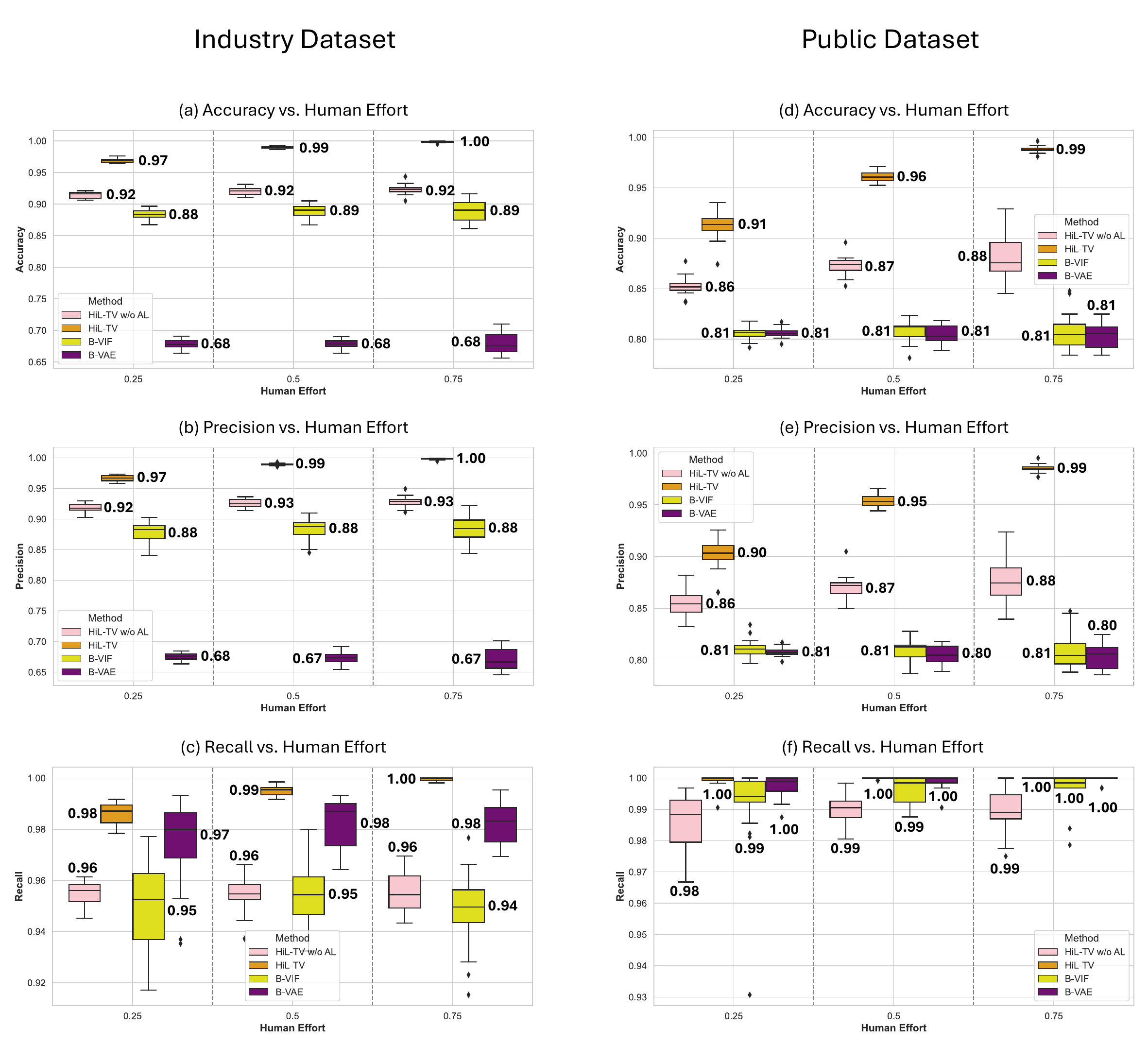}
    \vspace*{-0.4cm}
    \caption{Accuracy, precision, and recall distributions across 20 runs for \approach, \approach\ without active learning, and the two baselines, B-VIF and B-VAE, at varying levels of human effort for both the industry (left) and public (right) datasets.}
    \label{fig:RQ2_baselines}
    \vspace*{-0.4cm}
\end{figure*}

Figure~\ref{fig:RQ2_baselines} compares the accuracy, precision and recall distributions across the $20$ runs for each method and at different levels of human effort for both the industry and public datasets. As the figure shows, \approach\ leads to a significant increase in accuracy, precision, and recall compared to the baselines and \approach\ without active learning (i.e., \approach\ w/o AL in Figure~\ref{fig:RQ2_baselines}). Similarly, \approach\ w/o AL outperforms the baselines in accuracy and precision. For recall, \approach\ w/o AL is slightly worse than  the baselines in some experiments.  Table~\ref{tab:rq2_stats} shows Vargha-Delaney $\hat{A}_{12}$ statistic for accuracy, precision and recall, comparing  \approach\ and \approach\ w/o AL with each other and with the baselines. Note that all comparisons in Table~\ref{tab:rq2_stats} are statistically significant according to results from the Wilcoxon rank-sum test~\cite{capon:91}, yielding p-values~$\ll$~0.01. Hence, we show only the effect size values. 
According to Table~\ref{tab:rq2_stats}~(b), \approach\ significantly outperforms both baselines and \approach\ w/o AL in terms of accuracy and precision with large effect sizes. In terms of recall, \approach\ outperforms the baselines and \approach\ w/o AL with large effect sizes in 14 (out of 18) comparisons, with medium effect sizes in two comparisons, and with small effect sizes in the remaining two comparisons.  Similarly, \approach\ w/o AL significantly outperforms \emph{B-VIF} and \emph{B-VAE} in terms of accuracy and precision, with large effect sizes. Despite not outperforming the baselines in recall,  \approach\ w/o AL still achieves an average recall of at least 96\% on the industry dataset and at least 98\% on the public dataset.

\begin{table}[t]
\caption{The statistical test results for  accuracy, precision and
recall, comparing \approach\ and \approach\ w/o AL with each
other and with the baselines. All comparisons are statistically significant, yielding p-values $\ll$ 0.01. }~\label{tab:rq2_stats}
\vspace*{-0.3cm}
\centering

\begin{subtable}[t]{0.55\textwidth}
\caption{\approach\ w/o AL vs. baselines B-VAE and B-VIF}~\label{tab:RQ2_stat_hiltv_wo}
\scalebox{0.65}{
\hspace{1cm} 
\begin{tabular}{|c|c|ccc|ccc|}
\hline
\multirow{2}{*}{\textbf{Dataset}}  & \multirow{2}{*}{\textbf{\begin{tabular}[c]{@{}c@{}}Human\\  Effort\end{tabular}}} 
& \multicolumn{3}{c|}{\textbf{B-VAE}} & \multicolumn{3}{c|}{\textbf{B-VIF}} \\ \cline{3-8} 
                                   &                                        & \multicolumn{1}{c|}{\textbf{Accuracy}} & \multicolumn{1}{c|}{\textbf{Precision}} & \textbf{Recall} & \multicolumn{1}{c|}{\textbf{Accuracy}} & \multicolumn{1}{c|}{\textbf{Precision}} & \textbf{Recall} \\ \hline
\multirow{3}{*}{\textbf{Industry}} & \textbf{0.25}                          & \multicolumn{1}{c|}{1 (L)}             & \multicolumn{1}{c|}{1 (L)}              & 0.15 (L)        & \multicolumn{1}{c|}{1 (L)}             & \multicolumn{1}{c|}{1 (L)}              & 0.56 (S)        \\ \cline{2-8} 
                                   & \textbf{0.5}                           & \multicolumn{1}{c|}{1 (L)}             & \multicolumn{1}{c|}{1 (L)}              & 0 (L)           & \multicolumn{1}{c|}{1 (L)}             & \multicolumn{1}{c|}{1 (L)}              & 0.56 (S)        \\ \cline{2-8} 
                                   & \textbf{0.75}                          & \multicolumn{1}{c|}{1 (L)}             & \multicolumn{1}{c|}{1 (L)}              & 0 (L)           & \multicolumn{1}{c|}{0.98 (L)}          & \multicolumn{1}{c|}{0.98 (L)}           & 0.66 (M)        \\ \hline
\multirow{3}{*}{\textbf{Public}}   & \textbf{0.25}                          & \multicolumn{1}{c|}{1 (L)}             & \multicolumn{1}{c|}{1 (L)}              & 0.06 (L)        & \multicolumn{1}{c|}{1 (L)}             & \multicolumn{1}{c|}{1 (L)}              & 0.24 (L)        \\ \cline{2-8} 
                                   & \textbf{0.5}                           & \multicolumn{1}{c|}{1 (L)}             & \multicolumn{1}{c|}{1 (L)}              & 0.03 (L)        & \multicolumn{1}{c|}{1 (L)}             & \multicolumn{1}{c|}{1 (L)}              & 0.14 (L)        \\ \cline{2-8} 
                                   & \textbf{0.75}                          & \multicolumn{1}{c|}{1 (L)}             & \multicolumn{1}{c|}{1 (L)}              & 0.04 (L)        & \multicolumn{1}{c|}{1 (L)}             & \multicolumn{1}{c|}{0.99 (L)}           & 0.19 (L)        \\ \hline
\end{tabular}
}
\end{subtable}
\hfill
\vspace*{.2cm}
\begin{subtable}[t]{0.55\textwidth}
\caption{\approach\ vs. the three other test input validators}~\label{tab:RQ2_stat_hiltv}
\centering
\scalebox{0.55}{
\hspace{-2.4cm} 
\begin{tabular}{|c|c|ccc|ccc|ccc|}
\hline
\multirow{2}{*}{\textbf{Dataset}}  & \multirow{2}{*}{\textbf{\begin{tabular}[c]{@{}c@{}}Human\\  Effort\end{tabular}}} 
& \multicolumn{3}{c|}{\textbf{HiL-TV  w/o}}                                                          
& \multicolumn{3}{c|}{\textbf{B-VAE}}                                                                  
& \multicolumn{3}{c|}{\textbf{B-VIF}}                                                                  \\ \cline{3-11} 
                                   &                                        & \multicolumn{1}{c|}{\textbf{Accuracy}} & \multicolumn{1}{c|}{\textbf{Precision}} & \textbf{Recall} & \multicolumn{1}{c|}{\textbf{Accuracy}} & \multicolumn{1}{c|}{\textbf{Precision}} & \textbf{Recall} & \multicolumn{1}{c|}{\textbf{Accuracy}} & \multicolumn{1}{c|}{\textbf{Precision}} & \textbf{Recall} \\ \hline
\multirow{3}{*}{\textbf{Industry}} & \textbf{0.25}                          & \multicolumn{1}{c|}{1 (L)}             & \multicolumn{1}{c|}{1 (L)}              & 1 (L)           & \multicolumn{1}{c|}{1 (L)}             & \multicolumn{1}{c|}{1 (L)}              & 0.7 (L)         & \multicolumn{1}{c|}{1 (L)}             & \multicolumn{1}{c|}{1 (L)}              & 1 (L)           \\ \cline{2-11} 
                                   & \textbf{0.5}                           & \multicolumn{1}{c|}{1 (L)}             & \multicolumn{1}{c|}{1 (L)}              & 1 (L)           & \multicolumn{1}{c|}{1 (L)}             & \multicolumn{1}{c|}{1 (L)}              & 0.98 (L)        & \multicolumn{1}{c|}{1 (L)}             & \multicolumn{1}{c|}{1 (L)}              & 1 (L)           \\ \cline{2-11} 
                                   & \textbf{0.75}                          & \multicolumn{1}{c|}{1 (L)}             & \multicolumn{1}{c|}{1 (L)}              & 1 (L)           & \multicolumn{1}{c|}{1 (L)}             & \multicolumn{1}{c|}{1 (L)}              & 1 (L)           & \multicolumn{1}{c|}{1 (L)}             & \multicolumn{1}{c|}{1 (L)}              & 1 (L)           \\ \hline
\multirow{3}{*}{\textbf{Public}}   & \textbf{0.25}                          & \multicolumn{1}{c|}{1 (L)}             & \multicolumn{1}{c|}{0.99 (L)}           & 0.98 (L)        & \multicolumn{1}{c|}{1 (L)}             & \multicolumn{1}{c|}{1 (L)}              & 0.65 (M)        & \multicolumn{1}{c|}{1 (L)}             & \multicolumn{1}{c|}{1 (L)}              & 0.84 (L)        \\ \cline{2-11} 
                                   & \textbf{0.5}                           & \multicolumn{1}{c|}{1 (L)}             & \multicolumn{1}{c|}{1 (L)}              & 1 (L)           & \multicolumn{1}{c|}{1 (L)}             & \multicolumn{1}{c|}{1 (L)}              & 0.58 (S)        & \multicolumn{1}{c|}{1 (L)}             & \multicolumn{1}{c|}{1 (L)}              & 0.79 (L)        \\ \cline{2-11} 
                                   & \textbf{0.75}                          & \multicolumn{1}{c|}{1 (L)}             & \multicolumn{1}{c|}{1 (L)}              & 0.98 (L)        & \multicolumn{1}{c|}{1 (L)}             & \multicolumn{1}{c|}{1 (L)}              & 0.55 (S)        & \multicolumn{1}{c|}{1 (L)}             & \multicolumn{1}{c|}{1 (L)}              & 0.75 (L)        \\ \hline
\end{tabular}
}
\end{subtable}
\end{table}

\begin{tcolorbox}[breakable, colback=gray!10!white,colframe=black!75!black]
\textbf{RQ2 Answer:}
Both novel features of \approach, namely classification based on multiple image-comparison metrics and the inclusion of an active learning loop, lead to significant improvements in accuracy and precision. On average, the use of multiple metrics results in improvements of 4.6\% and 15\% in overall accuracy, and 5.4\% and 16\% in overall precision, respectively, compared to B-VIF and B-VAE. Furthermore, on average, the introduction of an active learning loop provides a 7.5\% improvement in accuracy, a 6.8\% improvement in precision, and a 2.1\% improvement in recall compared to not using active learning. Overall, \approach{} results in an average accuracy improvement of at least 12.9\% compared to the baselines.
\end{tcolorbox}

\begin{table*}[t]
\centering
\caption{Pearson correlation coefficients  between each metric in Table~\ref{tab:metrics_explained}  and the validity labels in the industry and public datasets. The values highlighted green represent the metrics with the highest correlations for each dataset. Bold values represent p-values $<0.05$, indicating high confidence in the correlations.  Gray values represent p-values larger than $0.1$.
}
\label{tab:metrics_correlation_label}
\scalebox{.9}{
\begin{tabular}{|c|c|c|c|c|c|c|c|c|c|c|c|c|c|}
  \hline
  Dataset & PSNR & CPL & CS & SSIM & MSE & WD & KL & SSS & TSI & $\mathit{Hist}\_{\mathit{cor}}$ & $\mathit{Hist}\_{\mathit{int}}$ & VAE-RE & VIF \\ 
  \hline
  Industry & \cellcolor{green} \textbf{0.47} & \cellcolor{green} \textbf{-0.58} & \cellcolor{green} \textbf{0.78} & \cellcolor{green} \textbf{0.54} & \textbf{-0.23} & \textbf{-0.26} & \textbf{-0.07} & \textbf{-0.36} & \textbf{-0.42} & \textbf{0.16} & \textbf{0.18} & \textcolor{gray}{0.02} & \cellcolor{green} \textbf{0.70} \\ 
  \hline
 Public & 0.03& \cellcolor{green} \textbf{-0.28}& 0.01& \cellcolor{green} \textbf{0.11}& \textcolor{gray}{0.04}& \textcolor{gray}{-0.04}& 0.01& \cellcolor{green} \textbf{-0.10}& -0.03& \cellcolor{green} \textbf{-017}& 0.01& \textbf{-0.09}& \cellcolor{green}\textbf{0.15}\\
  \hline
\end{tabular}
}
\vspace*{-.3cm}
\end{table*}

\subsection{RQ3 Experiments and Results}
\label{section:RQ3}
To address RQ3, we compute the Pearson correlation between each of the 13 metrics  in Table~\ref{tab:metrics_explained} and the validity labels in both our industry and public datasets. Table~\ref{tab:metrics_correlation_label} shows the Pearson correlation coefficients between each metric and the valid  labels for our two datasets. Note that correlations are generally higher for the industry dataset due to its significantly higher image resolutions compared to the public dataset (see Table~\ref{tab:dataset_info}).  The top five metrics with the highest absolute correlation coefficients  for each dataset are highlighted in Table~\ref{tab:metrics_correlation_label} and are as follows: PSNR, CPL, CS, SSIM, and VIF for the industry dataset, and CPL, SSIM, SSS, $\mathit{Hist}_{\mathit{cor}}$, and VIF for the public dataset. This shows that both feature-level and pixel-level metrics appear in the list of most influential metrics for both datasets. 
After identifying the top five metrics most correlated with the validity labels for each dataset, we calculate the pairwise Pearson correlations among them to evaluate potential redundancies. The results indicate that only CS and VIF have a strong correlation above 0.9 for the industry dataset. All other pairwise correlations for the industry dataset as well as all correlations for the public dataset are below 0.75, suggesting low redundancy. The Pearson correlation results for RQ3 are available in our replication package~\cite{github}.

\begin{tcolorbox}[breakable, colback=gray!10!white,colframe=black!75!black]
\textbf{RQ3 Answer:} Both feature-level and pixel-level \metrics\ are influential in validating test input images. Different \metrics, for the most part, contribute unique (non-redundant) information to the test input validation process.
\end{tcolorbox}
\vspace*{-1em}
\subsection{Threats to Validity}
\subsubsection{Internal validity} 
To prevent data leakage in RQ1 and RQ2, we used separate, non-overlapping training and test sets. In RQ2, to ensure a fair comparison between \approach\ and the two baselines, we used the same training and test sets for all methods. To counteract random variation in  RQ2, we randomly selected 20 subsets from our datasets for each effort level and repeated the experiments for each subset. Regarding the mitigation of bias in our datasets, we note the following: The ground-truth labels for the public dataset were taken directly from Hu et al.~\cite{marsha} (via majority voting), with no modifications made by us. For our industry dataset, labelling was performed by two independent human labellers who are not co-authors of this paper. The ground-truth labels were subsequently reviewed by domain experts as explained in Section~\ref{subsec:datasets}.

\subsubsection{External validity} 
The public dataset used in our evaluation is derived from CIFAR-10~\cite{Cifar10}, a well-known benchmark in the research community that has also been used previously for evaluating test input validators~\cite{marsha, WhenWhyTest2022, Dola_2021_DAIV}. The consistency observed between the results of our experiments on the public dataset and our domain-specific industry dataset provides a degree of confidence in the generalizability of our findings. To our knowledge, we are the first to evaluate and compare test input validation methods using an industry dataset. Nevertheless, as with any case study-based research, further evaluation across a broader range of study subjects is necessary to draw more definitive \hbox{conclusions about generalizability.}
\section{Related Work}
Ensuring the validity of test inputs is a critical aspect of software testing and remains an active area of research, particularly for DL systems, where testing methods are rapidly evolving~\cite{WhenWhyTest2022,Riccio_2020_SystematicMapping}.  Recent studies define invalid inputs as those that are under-represented in the training set of a given DL model and propose using out-of-distribution detection methods to identify such inputs~\cite{WhenWhyTest2022, Dola_2021_DAIV, Stocco_2020_selforacle}. Specifically, DAIV~\cite{Dola_2021_DAIV} trains a VAE and uses its loss function to discriminate valid and invalid test inputs. Self-oracle~\cite{Stocco_2020_selforacle} uses auto-encoders and time-series-based anomaly detection to discriminate valid and invalid test inputs for autonomous driving systems. Riccio and Tonella~\cite{WhenWhyTest2022} have shown that automated test input validators based on DAIV and self-oracle can achieve up to a 78\% agreement rate with human validators. Hu et al.~\cite{marsha} propose to discriminate between valid and invalid test inputs using  a metric based on VIF~\cite{vif}. 

Except for VIF and the VAE loss function (VAE-RE), the other eleven metrics in Table~\ref{tab:metrics_explained} have primarily been used to evaluate the effectiveness of image-to-image translators as a means of bridging the gap between simulated and real-world images for testing autonomous driving systems~\cite{stocco}. In our work, we use all the metrics  in Table~\ref{tab:metrics_explained}, including VIF and VAE-RE, to automate test input validation for image-based DL systems. We show that our approach, which combines multiple metrics with active learning, significantly outperforms baselines that rely on only VIF or VAE-RE.
\section{Lessons Learned}

\emph{Lesson 1: Validating test inputs using VAE  requires fine-tuning the architecture to suit the specific dataset.} The VAE-based test input validator performed comparably to the VIF-based method on the public dataset but significantly under-performed on the industry dataset compared to the other methods. The VAE architecture used in our experiments was adapted from prior studies~\cite{Amini_2024_Translators, WhenWhyTest2022}. Although, as described in RQ2, we trained VAE and independently optimized a threshold for the VAE-RE metric for each dataset, we applied the same VAE architecture across both datasets. The absence of dataset-specific architecture selection is likely the primary reason for VAE's poor performance on the industry dataset. This finding suggests the need for careful tuning of VAE's  architecture for each dataset. Given that practitioners may not always have the resources or expertise for such tuning, VAE-based test input validation may not be a universally practical technique.

\emph{Lesson 2: Test input validation methods that use multiple metrics outperform single-metric methods without requiring additional manual effort.} Our results show that using multiple \metrics\ provides better validation compared to using a single metric. A priori validation of a subset of the given dataset is necessary even for single-metric methods, as their thresholds do not generalize across different datasets. Therefore, using multiple metrics does not increase the manual effort compared to single-metric methods.

\emph{Lesson 3: Pixel-level and feature-level metrics are both useful, regardless of whether test inputs are generated through pixel-level or generative transformations.}  Our industry partner primarily uses generative models to create test inputs, whereas the test inputs from our public dataset are generated through pixel-level augmentations. In our experiments, both pixel-level and feature-level metrics turned out to be influential for both datasets. This suggests that incorporating both types of metrics, as we do in \approach, is beneficial regardless of the method used to generate the test inputs.

\emph{Lesson 4: Active learning improves input validation accuracy without adding to human effort.} Active learning identifies inputs that are challenging for the automated classifier and requests human labelling for those specific cases. This, in turn, allows human validators to focus on more difficult cases, thereby improving overall accuracy while reducing human involvement in simpler cases. Our experiments indicate that active learning significantly increases validation accuracy for test inputs without requiring additional effort compared to methods that do not use active learning.

Our \textbf{replication package} including our code, our test input validation tool and our public dataset is available online~\cite{github}.

\section*{Acknowledgements}
We gratefully acknowledge funding from Mitacs Accelerate,
SmartInsideAI, and NSERC of Canada under the Discovery and Discovery
Accelerator programs.

\bibliographystyle{plain}
\bibliography{references}

\end{document}